\documentclass[usenatbib]{mn2e}
\usepackage{graphicx}

\title[Spectral evolution of nova V1494 Aql]{Optical and Near-Infrared spectroscopy of Nova V1494 Aquilae 1999 \# 2}

\author[Kamath, U. S. et al.]{U. S. Kamath,$^{1,2}$ 
\thanks{E-mail : kamath@crest.ernet.in (USK) ; gca@iiap.res.in (GCA) ; ashok@prl.ernet.in (NMA) ; ydm@inaoep.mx (YDM) ; dks@crest.ernet.in (DKS)}
G. C. Anupama,$^{2}$ N. M. Ashok,$^{3}$ Y. D. Mayya $^{4}$ and D. K. Sahu$^{1,2}$ \\
$^{1}$ Centre for Research and Education in Science and Technology, Hosakote 562114, India \\ 
$^{2}$ Indian Institute of Astrophysics, Koramangala, Bangalore 560 034, India \\
$^{3}$ Physical Research Laboratory, Navrangpura, Ahmedabad 380 009, India \\
${4}$ Instituto Nacional de Astrofisics Optica y Electronica, Apdo Postal 51 y 216, 72000 Puebla, Pue., Mexico }

\begin{document}

\date{Received / Accepted}

\pagerange{\pageref{firstpage}--\pageref{lastpage}} \pubyear{2005}

\maketitle

\label{firstpage}

\begin{abstract}

Optical and near-infrared spectroscopic observations of the fast nova V1494 Aquilae 1999 \# 2 covering various phases -- early decline, transition and nebular -- during the first eighteen months of its post-outburst evolution are presented in this paper. During this period, the nova evolved in the  P$_{\rm fe}$P$_{\rm fe}^{\rm o}$C$_{\rm o}$ spectral sequence. The transition from an optically thick wind to a polar blob - equatorial ring geometry is seen in the evolution of the spectral line profiles. There is evidence of density and temperature
stratification in the ejecta. Physical conditions in the ejecta have been estimated based on our observations.

\end{abstract}


\section{Introduction}
\label{intro}

Nova V1494 Aquilae 1999 \# 2 was discovered by Pereira on Dec 1.875, 1999 at a visual magnitude of 6.0 \citep{pereira}. Spectroscopic observations  showed 
emission lines of the Balmer series, \mbox{O\,{\sc i}}, \mbox{Mg\, {\sc ii}}, 
\mbox{Fe\,{\sc ii}}, all having P-Cyg profiles (\citealt{fujii} ; 
\citealt{ayani} ; \citealt{moro}). This confirmed the object to be an ``Fe-II'' nova in its early stage. The nova reached a maximum brightness of 4.0 on 
December 3.4, 1999, followed by a rapid decline with characteristic timescales 
of $t_2=6.6\pm 0.5$~days and $t_3=16\pm 0.5$~days, making this a fast nova 
\citep{kt}. The light curve showed oscillations during the  transition phase, 
between mid-January to April 2000, which was followed by a smooth decline to 
15 mag \citep{kky}.  Early spectral evolution in the optical was reported by \cite{kt} and \cite{gca}. Spectral evolution in the optical indicates 
likely continued mass ejection for over 195 days (\citealt{ie} ; \citealt{eyres}). The interstellar extinction to the nova has been estimated to be E(B-V) = 0.6 from the equivalent widths of the interstellar absorption components of \mbox{Na\,{\sc i}} D1 and D2 \citep{ie}. Using this result, the authors have estimated the distance to the nova as 1.6 kpc. We shall adopt all the above parameters in our discussion and calculations (see Sec. \ref{abundances}).

In this paper we present multi-epoch optical and near-infrared spectra of  nova V1494 Aquilae obtained over a period of 18 months since outburst at various  phases of its evolution.

\section{Observations}

\subsection{Optical}
Optical CCD spectra were obtained from the Vainu Bappu Observatory on several
nights in December 1999 using the OMR spectrograph at the cassegrain focus of
the 2.3m Vainu Bappu Telescope (VBT). The spectra were obtained at a resolution of 11~\AA. FeAr and FeNe
spectra were used for wavelength calibration. 58 Aql was used as the spectrophotometric standard
star \citep{hamuy}. Spectra were also obtained from the Guillermo Haro Astrophysical Observatory (GHAO) during
December 4-7 using the B\&C spectrograph on the 2.12m telescope, at 
resolutions of 10~\AA\, and 2~\AA. The nova was observed during the nebular phase
on two occasions in 2001 (April 29 and May 9) from VBT. All spectra were bias subtracted and flat-field
corrected in the standard manner, and the one dimensional spectra extracted
using the optimal extraction method. The spectra were wavelength calibrated 
 and corrected for instrumental response in the standard manner.  All data were
analysed using various tasks within IRAF. 

The spectra are not corrected for telluric absorption. These features are marked in Fig. \ref{gho}. The absolute flux level is accurate to 10\%. On a relative scale, the emission line fluxes are generally accurate to 10\%. However, weaker lines have errors upto 20\% while strong lines like H$\beta$ are accurate to 5\%.

\subsection{Near-Infrared}
Near-infrared spectra covering the early decline and transition phases were obtained with PRLNIC, an imager-spectrometer based on a 256 $\times$ 256 HgCdTe NICMOS3 array detector, at the cassegrain focus of the 1.2m telescope of Mt. Abu IR Observatory (MAIRO). To remove the background, spatially offset spectra were obtained by nodding the star along the slit oriented in the north-south direction. One or more spectra were obtained at each position. Nearby standard stars were observed in the same manner.  HR 7953, $\beta$ Tau, $\delta$ Aql, HR 5407, $\lambda$ Ser and $\alpha$ Leo were used as standard stars on different nights. Wavelength calibration was done using the OH sky lines. Known stellar absorption features were removed from the standard star spectra and the nova spectrum was then divided by this. The result was multiplied by the spectrum of a blackbody of temperature commensurate with the spectral type of the standard star. This process removes the effects of atmospheric absorption and intrumental response in the wavelength regions common to the nova and standard star spectra; the nova spectra have slightly different spectral coverage at different epochs.  The data were reduced using standard procedures available within IRAF. The fluxes are on a relative scale and accurate to around 10\%.

Table \ref{obs_log} gives the details of both optical and near-infrared observations.

\begin{table*}
\begin{minipage}{126mm}
\caption{Log of observations.}
\label{obs_log}
\centering
\begin{tabular}{l l l l l}
\hline\hline
Date & Coverage &  Resolution & Observatory  & Instrument  \\
\hline
\multicolumn{5}{l}{Optical spectra} \\

4 Dec 1999 & 4000 -- 9000 \AA & 10 \AA & GHAO & B\&C\\
5 Dec 1999 & 4000 -- 9000 \AA & 2 \AA & GHAO & B\&C\\
6--7 Dec 1999 & 4000 -- 9000 \AA & 10 \AA & GHAO & B\&C\\
6--7 Dec 1999 & 3700 -- 9500 \AA & 11 \AA & VBT & OMR \\
9--10 Dec 1999 & 4000 -- 9500 \AA & 11 \AA & VBT & OMR \\
15--16 Dec 1999 & 4000 -- 9500 \AA & 11 \AA & VBT & OMR \\
23 Dec 1999 &  4500 -- 7000 \AA & 11 \AA & VBT & OMR \\
25 Dec 1999 &  4000 -- 9000 \AA & 11 \AA & VBT & OMR \\
29 Apr 2001 &  4000 -- 8000 \AA & 11 \AA & VBT & OMR \\
9 May 2001 &  4800 -- 9400 \AA & 11 \AA & VBT & OMR \\
\\
\multicolumn{5}{l}{Near-Infrared spectra}\\
5 Dec 1999 & 1 -- 2.5 $\mu$m & 2 $\times$ 10$^{-3}$ $\,\mu$m & MAIRO & PRLNIC\\
7 Dec 1999 & 1 -- 2.5 $\mu$m & 2 $\times$ 10$^{-3}$ $\,\mu$m & MAIRO & PRLNIC\\
31 Dec 1999 & 1 -- 2.5 $\mu$m & 2 $\times$ 10$^{-3}$ $\,\mu$m & MAIRO & PRLNIC\\11 Feb 2000 & 1 -- 2.5 $\mu$m & 2 $\times$ 10$^{-3}$ $\,\mu$m & MAIRO & PRLNIC\\1 Mar 2000 & 1 -- 2.5 $\mu$m & 2 $\times$ 10$^{-3}$ $\,\mu$m & MAIRO & PRLNIC \\\hline
\end{tabular}
\end{minipage}
\end{table*}

\section{Evolution of the spectra}

Nova spectra undergo different phases of evolution depending on the physical conditions of the outflowing gas and temperature of the central remnant. A nova is initially seen as an optically-thick fireball, with numerous absorption lines and hardly any emission lines. As the ejecta expands outwards and becomes optically thin with time numerous emission lines begin to appear. A high-ionisation coronal phase is reached in some novae. Eventually, the nova returns to its pre-outburst state.

\subsection{Fireball phase}
Our first spectrum was obtained on 4 December 1999, at near-maximum.  This spectrum (Fig.\ref{gho} -- top) displays the fireball phase of the nova explosion. Many members of the Hydrogen Balmer series, right upto Balmer 21, can be seen along with several lines of \mbox{O\,{\sc i}}. Signature of the outflowing optically-thick wind can be seen as P-Cyg profiles on all these lines. Numerous lines of \mbox{Fe\,{\sc ii}}, \mbox{N\,{\sc i}} , \mbox{Mg\,{\sc i}} and \mbox{C\,{\sc i}}
are also present, although the P-Cyg profiles of some of them are not apparent due
to blending with adjacent lines. The absorption components of some lines like the \mbox{Ca\,{\sc ii}} IR triplet are much stronger than the emission.  The absorption velocites lie in the range of 1000-1500 km s$^{-1}$. Thus, the spectrum is that of optically thick  gas  flowing outwards from the central source. There possibly  are some warm, dense packets of gas in the outflowing material.

\subsection{Early decline and transition phase}

\begin{figure}
\resizebox{\hsize}{!}{\includegraphics{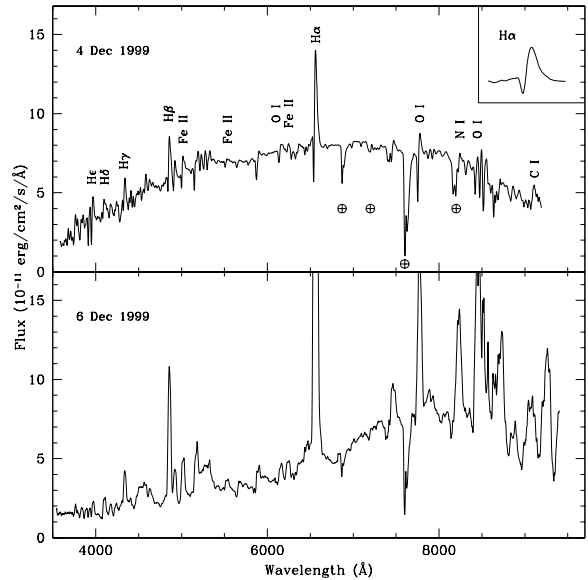}}
\caption[]{(Top) Spectrum of V1494 Aql in the fireball (pseudophotosphere expansion) phase. Several absoprtion lines can be seen and all the emission lines display strong P-Cygni profiles. Terrestrial absorption lines are marked. Inset shows the profile of the H$\alpha$ line. Prominent lines are labelled. (Bottom) Spectrum of V1494 Aql  displaying the evolution from optically thick to partially optically thin phase. The emission lines have strenghtened  and the P-Cygni profiles have weakened. Notice the change in the H$\alpha$ profile.}
\label{gho}
\end{figure}

In the early decline phase of novae, the P-Cygni profiles disappear and the lines acquire a more rounded emission peak 
 This transition can be seen in our spectra obtained on 6 December 1999 (Fig. \ref{gho}-- bottom). Polarisation observations by \citet{kawabata} during this period revealed a drastic change   in the position angle of the intrinsic polarisation (from 65\degr to 140\degr). They have interpreted this in terms of an optical depth effect or a geometric change in the nova wind. A combination of both these effects are likely to be responsible for this change.

\begin{figure}
\resizebox{\hsize}{!}{\includegraphics{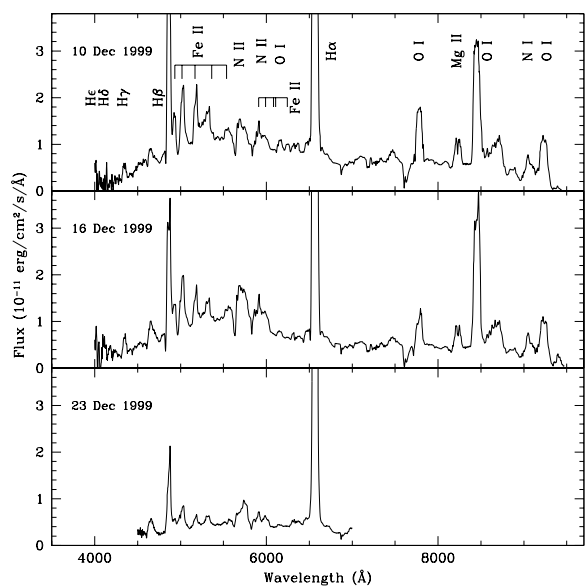}}
\caption[]{Spectra of V1494 Aql in the early decline phase. The gradual hardening of the spectrum is reflected in the weakening of \mbox{Fe\,{\sc ii}} lines and the sustained strength of lines of higher ionisation.}
\label{vbt}
\end{figure}

\begin{figure}
\resizebox{\hsize}{!}{\includegraphics{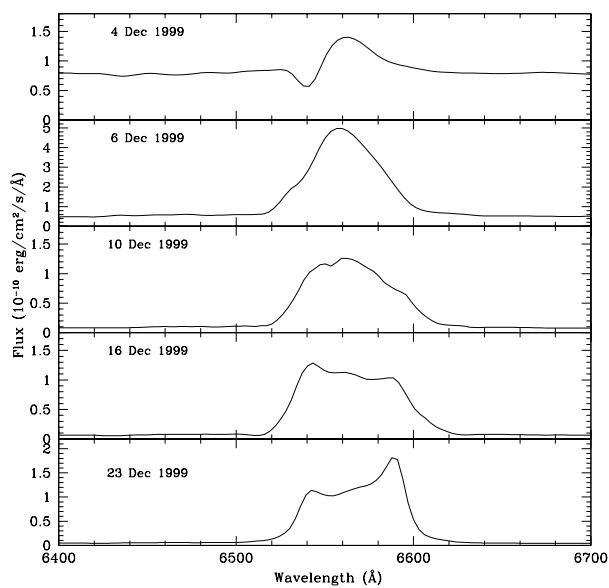}}
\caption[]{Evolution of the H$\alpha$ line showing the transition from an optically thick wind to a polar blob -- equatorial ring geometry.}
\label{halpha}
\end{figure}

Subsequent optical spectra displayed in Fig. \ref{vbt} show typical 
characteristics of Fe-II novae -- numerous \mbox{Fe\,{\sc ii}} and  
\mbox{N\,{\sc ii}} lines -- along with prominent lines of H$\alpha$, H$\beta$ 
and \mbox{O\,{\sc i}}. The emission lines show P-Cyg profiles on the first few 
days and  become more rounded subsequently. An optically thin wind gives a rounded emission peak, while an optically thin shell produces a more rectangular line profile (\citealt{williams} and references therein). The profile of the H$\alpha$ line shown in Fig. \ref{halpha} clearly demonstrates the prominence of the wind and shell components at different times. Initially, it shows a strong P-Cyg absorption and is representative of an optically-thick wind. Later it is characteristic of optically thin wind and eventually shows the typical saddle shape of the polar blob - equatorial ring geometry of the ejected shell (see Sec. \ref{geometry}). The hardening of the spectrum with time can be seen in the weakening of the \mbox {Fe\,{\sc ii}} and \mbox{N\, {\sc ii}} lines, and the continuing 
strength of the \mbox{O\,{\sc i}} 8446 \AA\, line. Some lines like 
\mbox{He\,{\sc i}} (4471~\AA\,, 5876~\AA\,) and \mbox{[O\,{\sc i}]} (6300, 
6363~\AA\,) are clearly seen in emission only on some days. The FWHM of the H$\beta$ line lies in the range of 2280 -- 2875 km s$^{-1}$ in this period, with no apparent secular change.

\begin{table*}
\caption{Line identification and observed fluxes relative to H${\beta}$ in the early decline phase. Fluxes with higher errors are marked with a colon. See the text for a discussion of the errors.}
\label{early_fluxes}
\centering
\begin{tabular}{l l r r r r r}
\hline\hline
$\lambda$ (\AA) & Line id & \multicolumn{5}{c}{December 1999}\\
& & 6 &  7 &  10 & 16 &  23\\
\hline
3770 & Balmer 11 & 0.02: & 0.01: & & & \\
3835 & Balmer 9 & 0.06 & 0.02: & & & \\
3889 & Balmer 8 & 0.05 & 0.03: & & & \\
3933 & Ca {\sc ii} K(1) & 0.04 & 0.02: & & & \\
3970 & Ca {\sc ii} H(1) + H${\epsilon}$ & 0.10 & 0.06 & & & \\
4101 & H${\delta}$ & 0.14 & 0.09 &  & 0.14 &  \\
4173 & Fe {\sc ii}(27)& 0.01: & 0.06 & & 0.07 & \\
4233 & Fe {\sc ii} & & 0.08 & & 0.05 & \\
4340 & H${\gamma}$ & 0.22 & 0.16 & 0.03: & 0.13 & \\
4351 & Fe {\sc ii} (27) & & 0.03: & & \\
4471 & He {\sc i} ? & 0.02: & 0.04 & & & \\
4556 & Fe {\sc ii} & 0.10 & 0.09 & & & \\
4584 & Fe {\sc ii}(38) & 0.08 & 0.07 & & & \\
4635 & Fe {\sc ii} & 0.09 & 0.09 & 0.03: & & \\
4649 & O {\sc ii} & 0.04 & 0.05 & 0.03: & 0.15 & 0.21 \\
4861 & H${\beta}$ & 1.00 & 1.00 & 1.00 & 1.00 & 1.00\\
4924 & Fe {\sc ii}(42) & 0.27 & 0.21 & 0.18 & 0.19 & 0.06 \\
5018 & Fe {\sc ii} & 0.35 & 0.33 & & & \\
5169 & Fe {\sc ii} + Mg {\sc i} & 0.12 & 0.56 & 0.09 & 0.10 & \\
5176 & N {\sc ii} + Mg {\sc i} &  & &  &  & 0.17 \\
5184 & Mg {\sc i} & 0.24 & & 0.13 & 0.13 & \\
5275 & Fe {\sc ii} (49) + (84) & & & 0.05 & & \\
5316 & Fe {\sc ii} (49) & & & 0.01: & & \\
5363 & Fe {\sc ii}(48) & 0.10 & & 0.04 & 0.03: & \\
5425 & Fe {\sc ii} & 0.04 & 0.03: & & & \\
5535 & Fe {\sc ii}(55) + N {\sc ii} & 0.10 & 0.10 & & & \\
5680 & N {\sc ii}(3) & 0.08 & 0.08 & 0.15 & & 0.30\\
5755 & [N {\sc ii}](3) & 0.05 & 0.07 & 0.10 & & 0.66\\
5876 & He {\sc i} ? & 0.04 & & & & \\
5909 & Fe {\sc ii} & 0.12 & & 0.14 & 0.18 & 0.23 \\
5942 & N {\sc ii}(28) & 0.07 & 0.13 & & & \\
5991 & Fe {\sc ii}(46) & 0.04 & 0.04 & 0.12 & & 0.28\\
6084 & Fe {\sc ii}(46) & 0.03: & & & & \\
6157 & O {\sc i} & 0.15 & 0.14 & 0.08 & 0.06 & 0.06 \\
6243 & Fe {\sc ii} + N {\sc ii} & 0.17 & 0.17 & 0.05 & & \\
6300 & [O {\sc i}] & 0.03: & & 0.04  & & 0.10\\
6363 & [O {\sc i}] & & & 0.04  & & 0.08\\
6563 & H$\alpha$ & 5.92 & 8.0 & 3.55 & 5.86 & 13.2 \\
7774 & O {\sc i} & 1.53 & 2.87 & 0.47 & 0.37 & \\
8232 & O {\sc i} (34) + Mg {\sc ii}(7) & 1.34 & 2.00 & 0.24 &  0.29 & \\
8446 & O {\sc i} & 2.57 & 4.63 & 1.18 & 1.82 & \\
8662 & Ca {\sc ii} (2) + Pa 13 & & 2.26 & & & \\
8863 & Paschen 11 & & 0.87 & & &  \\
9042 & N {\sc i}(15) & 0.45 & 0.75 & 0.10 & 0.12 & \\
9112 & C {\sc i} & 0.27 & & & & \\
9229 & Paschen 9 & & & 0.33 & 0.44 & \\
9264 & O {\sc i} & 1.23 & 2.55 & & & \\
9405 & C {\sc i} (9) & & 0.84 & & &  \\
9546 & Pa$\epsilon$ & & 0.52 & & &  \\
\\
\multicolumn{7}{c}{F$_{{\rm H}\,\beta}$ (10$^{-9}$ ergs/cm$^{2}$/s)}\\
& & 3.55 & 2.88 & 1.96 & 1.37 & 0.65 \\
\hline
\end{tabular}
\end{table*}

\begin{figure}
\resizebox{\hsize}{!}{\includegraphics{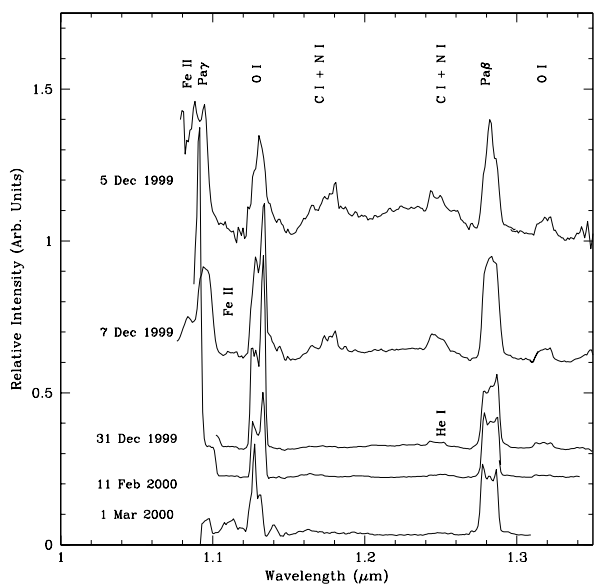}}
\caption[]{$J$ band spectra of V1494 Aql in the early decline and transition phases. Spectra obtained on different nights have been scaled and shifted for clarity. Increasing ionisation level of the ejecta is reflected in the gradual weakening of carbon and nitrogen lines and appearance of helium lines.}
\label{jspec}
\end{figure}

\begin{figure}
\resizebox{\hsize}{!}{\includegraphics{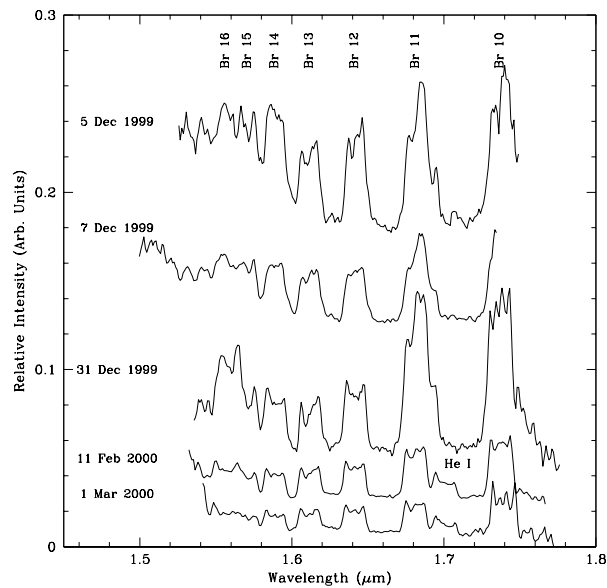}}
\caption[]{$H$ band spectra of V1494 Aql in the early decline and transition phases. This spectral region mainly covers the hydrogen Brackett series, but some unidentified lines (see text for details) and  the He I 1.700 $\mu$m line can also be seen.}
\label{hspec}
\end{figure}

\begin{figure}
\resizebox{\hsize}{!}{\includegraphics{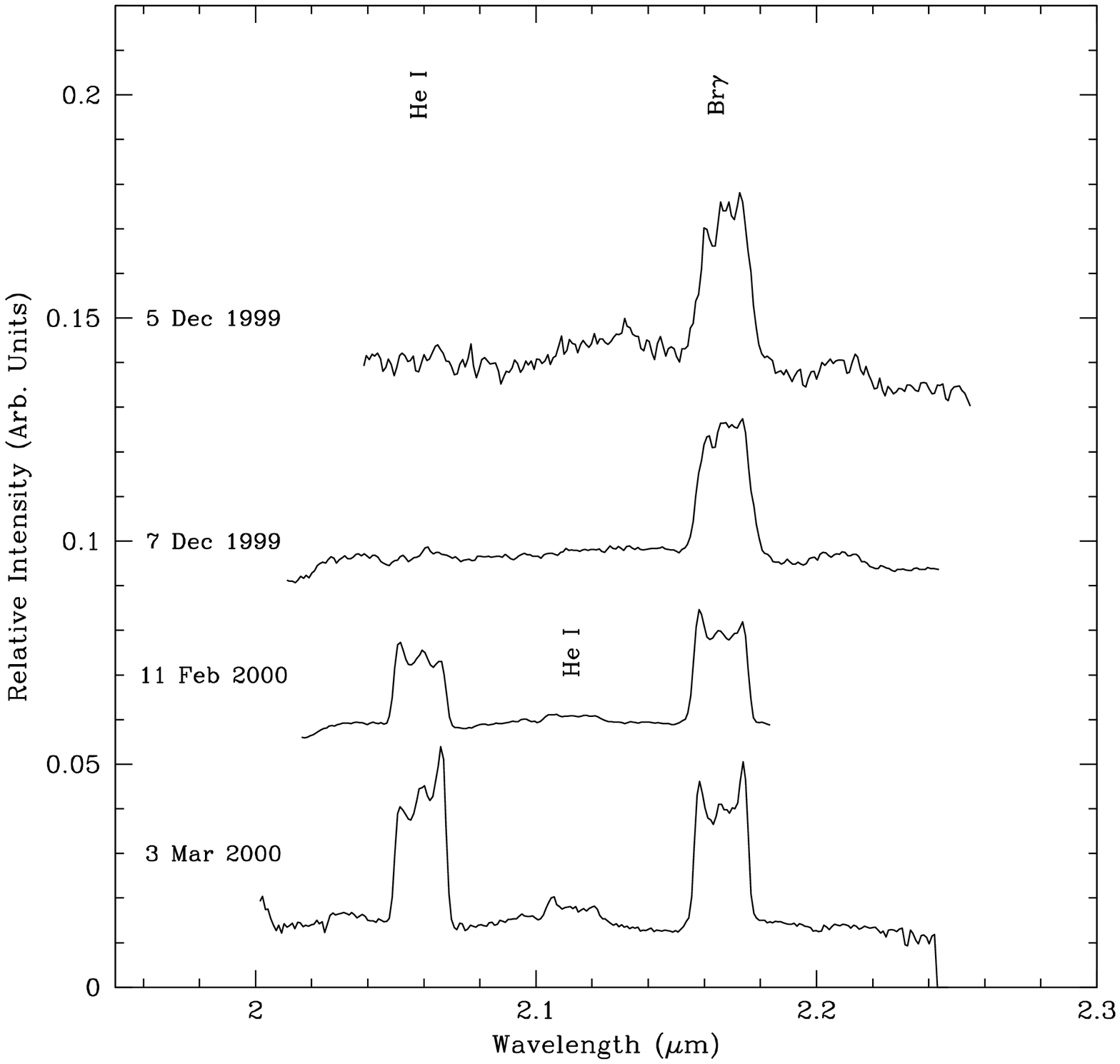}}
\caption[]{$K$ band spectra of V1494 Aql. The increasing level of ionisation is apparent in the emergence and strengthening of the helium lines. }
\label{kspec}
\end{figure}

Near-infrared spectra obtained on various days are shown in the figures \ref{jspec}, \ref{hspec} and \ref{kspec}. In each of these panels, the spectra have been offset from each other and scaled for illustrative purposes.
Lines of hydrogen (Paschen and Brackett series) and oxygen are prominent in the 5 December 1999 spectra. The $J$ band has several lines due to \mbox{C\,{\sc i}} and \mbox{N\,{\sc i}}. The overall spectrum is typical of an Fe-II nova in the very early decline stage. None of the lines show significant P-Cygni profiles, sugesting that the nova ejecta had become optically thin in the near-infrared as early as day 2. The increasing level of ionisation of the ejecta is reflected in the gradual decrease in the line strengths of \mbox{Fe\,{\sc ii}}, \mbox{C\,{\sc i}} and \mbox{N\,{\sc i}}, and the appearance of various lines of \mbox{He\,{\sc i}}. The strong line near 1.08 $\mu$m seen on 5 December 1999 is likely to be \mbox{Fe\,{\sc ii}}, as is corroborated by its reduced strength two days later.  The \mbox{He\,{\sc i}} lines at 1.252, 1.700 and 2.058 $\mu$m are seen starting from  respectively 28, 70 and 70 days since maximum.  
The \mbox{He\,{\sc i}} lines are very strong in March 2000. The FWHMs of the hydrogen and helium lines lie in the range 2100 -- 2800 km~s$^{-1}$. Again, no temporal trend can be seen.

\subsection{Late stage nebular spectra}
\label{nebular}

\begin{figure}
\resizebox{\hsize}{!}{\includegraphics{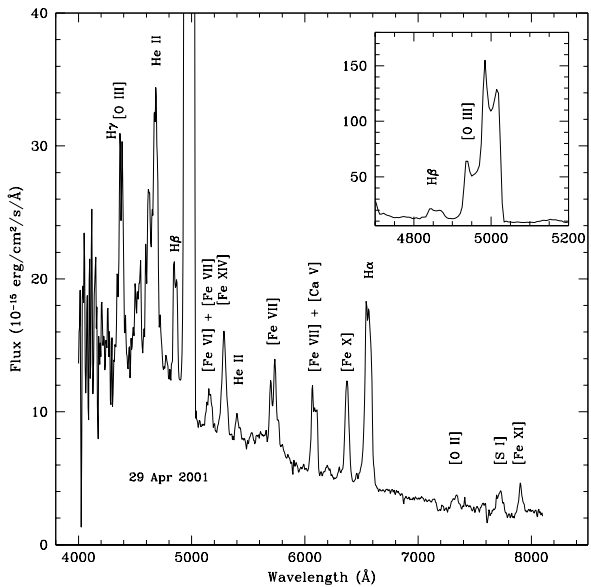}}
\caption[]{Spectrum of V1494 Aql in the coronal phase obtained in April 2001. Many  lines  of highly ionised iron can be seen. Inset shows the region around the \mbox{[O\,{\sc iii}]}  lines.}
\label{coronal_apr01}
\end{figure}

\begin{figure}
\resizebox{\hsize}{!}{\includegraphics{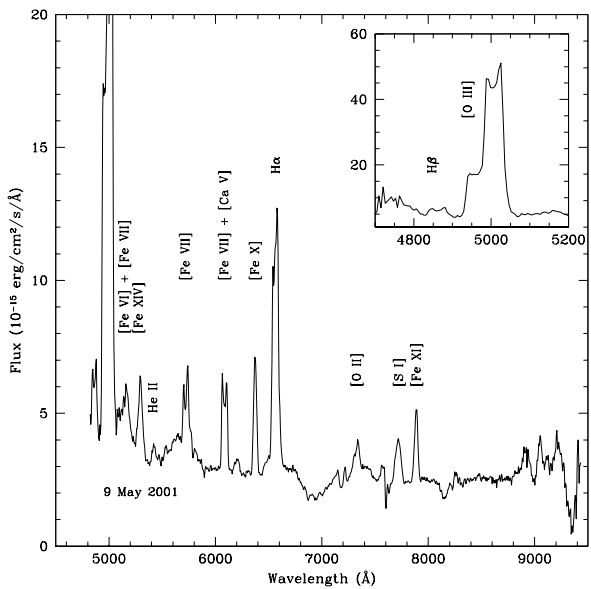}}
\caption[]{Spectrum of V1494 Aql in the coronal phase obtained in May 2001. The emission lines have strengthed further showing a continued hardening of the spectrum. Inset shows the region around the \mbox{[O\,{\sc iii}]} lines.}
\label{coronal_may01}
\end{figure}

V1494 Aql showed fairly strong nebular lines and weak coronal lines at around 65 and 80 days respectively \citep{ie}, during the transition phase. Infrared coronal lines seen in July 2000 (day 226), along with lines of \mbox{H\,{\sc i}}, \mbox{He\,{\sc i}} and \mbox{He\,{\sc ii}} \citep{venturini} show that the nova ejecta had zones of low or no ionisation coeval with the highly ionised zone. Optical spectra (\citealt{arkhipova} ; \citealt{ie}) show that by September 2000 (day $\sim$ 280) the coronal lines had strenghtened considerably while the low ionisation lines, such as \mbox{He\,{\sc i}}, had weakened significantly. Our spectra (see Figs. \ref{coronal_apr01} and \ref{coronal_may01}) show no  \mbox{He\,{\sc i}} lines, indicating that the helium has been completely ionised by day $\sim$ 510. Similarly, the absence of \mbox{[O\,{\sc i}]} 6300 \AA\, line implies that the zone where neutral oxygen was present earlier also has been ionised. Both spectra show many lines of highly ionised iron.

{\it CHANDRA} observations show that the nova had become a super-soft X-ray source by August 2000 \citep{starrfield}.  This is the signature of the energy emitted by hydrostatic hydrogen burning (of the unejected matter) on the surface of the white dwarf.  The April 2001 spectrum (see figure \ref{coronal_apr01}), shows  that the lines of highly ionised iron have sharp, single-peaked profiles. The nebular lines are double peaked and show similar structure as that of the Balmer lines. This suggests that the coronal lines could be arising in a region different (possibly closer to the hot white dwarf) from that of the nebular (and other) lines. This conforms to the model of a nova shell presented by \cite{saizar} wherein the ejecta consist of warm (T $\sim\,10^{4}$ K), dense clouds embedded in a hot (T $\sim\,10^{6}$ K), tenuous gas. The two phases are photoionised by the hard white dwarf continuum and the warm phase is further photoionised by the free-free continuum generated in the hot gas. Thus, the nebular spectra require some kind of non-uniform (in temperature and density) shell structure in order to explain the observed range of ionisation. 

In section \ref{abundances} below we have derived values for the physical conditions of the ejecta in this stage.

\begin{table}
\caption{Line identification and observed fluxes relative to H${\beta}$ in the  coronal phase. The fluxes are accurate up to 10\%.}
\label{coronal_fluxes}
\centering
\begin{tabular}{c l r r }
\hline\hline
$\lambda$ (\AA)& Line id & \multicolumn{2}{c}{2001}  \\
& & April & May \\
\hline
4340 + 4363 & H${\gamma}$ + \mbox{[O\,{\sc iii}]} & 1.90 & \\
4686 & \mbox{He\,{\sc ii}} & 2.96 & \\
4861 & H${\beta}$ & 1.00 & 1.00 \\
4959 + 5007 & \mbox{[O\,{\sc iii}]} & 23.06 & 20.49 \\
5159 & \mbox{[Fe\,{\sc vi}]} + \mbox{[Fe\,{\sc vii}]} & 0.31 & \\
5303 & \mbox{[Fe\,{\sc xiv}]} & 1.04 & 0.92 \\
5411 & \mbox{He\,{\sc ii}} & 0.22 & \\
5720 & \mbox{[Fe\,{\sc vii}]} & 0.95 & 1.30 \\
6087 & \mbox{[Fe\,{\sc vii}]} + \mbox[{Ca\, {\sc v}]} & 1.0 & 1.23 \\
6310 & \mbox{[S\,{\sc iii}]} & 0.05 & 0.24 \\
6374 & \mbox{[Fe\,{\sc x}]} & 0.92 & 1.11 \\
6563 & H${\alpha}$ & 2.68 & 4.02 \\
7330 & \mbox{[O\,{\sc ii}]} & 0.17 & 0.43  \\
7727 & \mbox{[S\,{\sc i}]}  & 0.25 & 0.69 \\
7891 & \mbox{[Fe\,{\sc xi}]} & 0.25 & 0.75 \\
9069 & \mbox{[S\,{\sc iii}]} & & 0.21 \\
\\
\multicolumn{4}{c}{F$_{{\rm H}\,\beta}$ (10$^{-13}$ ergs/cm$^{2}$/s)}\\
& & 3.44 & 1.41 \\
\hline
\end{tabular}
\end{table}

\subsection{Spectral evolutionary sequence}

The Tololo classification system for novae has been evolved to define the temporal evolution of a nova spectrum (\citealt{ctio1}, \citealt{ctio2}). Every nova can be assigned an evolutionary sequence according to the various phases and sub-classes for each phase that are observed in the optical spectra.
In the early decline phase (early-December 1999), permitted lines of Fe II were 
the strongest non-Balmer lines ; hence, the spectral classification of the 
nova during this phase is P$_{\rm fe}$. Also, the 
\mbox{O\,{\sc i}} 8446~\AA\, line remained stronger than H${\beta}$ while the 
nova was still in the permitted line phase. Hence a classification of 
P$_{\rm fe}^{\rm o}$ can be assigned. If \mbox{[Fe\,{\sc x}]} 6375~\AA\, 
emission is clearly present and stronger than \mbox{[Fe\,{\sc  vii}]} 6087~\AA, 
the nova spectrum is considered to be in the coronal phase, regardless of any 
other line strengths. This is the case during April 2001. The strongest 
non-Balmer line in this spectrum is \mbox{[O\,{\sc iii}]} 5007~\AA, and 
therefore the nova is in the C$_{\rm o}$ phase. Thus, the evolutionary sequence for V1494 Aql is P$_{\rm fe}$P$_{\rm fe}^{\rm o}$C$_{\rm o}$.

\section{Discussion}

\subsection{Ejecta geometry}
\label{geometry}

\begin{figure}
\resizebox{\hsize}{!}{\includegraphics{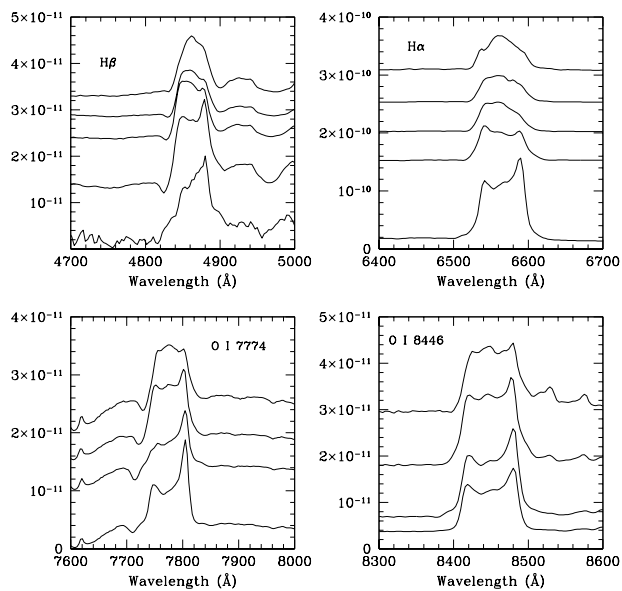}}
\caption[]{Temporal evolution of some optical emission line profiles in spectra obtained during early- (top) and late-December 1999 (bottom). The spectra have been shifted or scaled for illustrative purposes.}
\label{profiles_opt}
\end{figure}

\begin{figure}
\resizebox{\hsize}{!}{\includegraphics{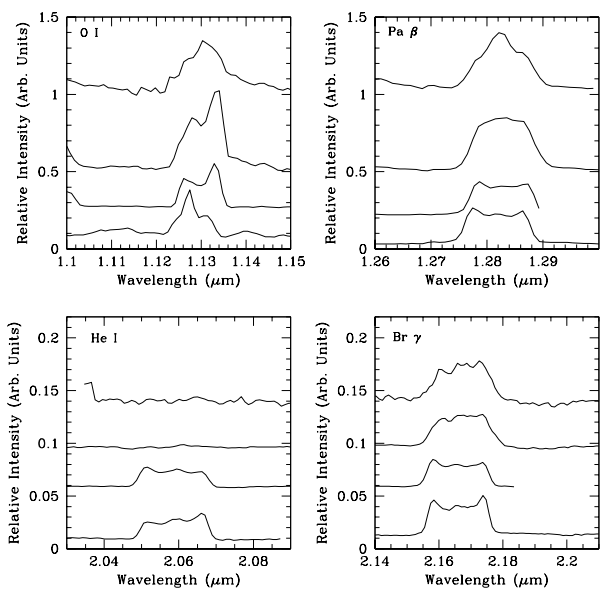}}
\caption[]{Evolution of some near-infrared emission line profiles seen in the early (December 1999 -- top) to the later (March 2000 -- bottom) spectra.The spectra have been shifted or scaled for illustrative purposes.}
\label{profiles_ir}
\end{figure}

Nova shells are not uniform, but show considerable structure because of the clumpy nature of the ejecta (for example, \citealt{anupama}). A double-peaked, saddle-shaped emission profile is the signature of emission from a shell of material, which has an equatorial-ring, polar-cone/blob morphology \citep{hutchings}. Substructures within the shell manifest themselves as multiply-peaked spectral lines \citep{gill}. Nova V1494 Aql showed triangular lines in the initial phases (see figures \ref{profiles_opt} and \ref{profiles_ir}). They broadened and acquired a more rectangular, multi-peaked  profile with time. The velocities deduced from emission peaks range from 500 to 1200 kms$^{-1}$, although the \mbox{O\,{\sc i}} lines show higher velocity components of upto 2500 km s$^{-1}$. Spectropolarimetric observations of this nova  have shown that an asymmetric geometry was present even prior to maximum brightness \citep{kawabata}. At around 10 days after maximum light, rapidly variable components of polarisation were observed, and their contribution increased with time. This is due to clumping in ejection near the nova. A radio image obtained using MERLIN on day 136 shows that the ejecta continued to have this clumpy structure \citep{eyres}.

The line profiles (Figures \ref{profiles_opt}, \ref{profiles_ir}) seen in the later decline phases are similar to those seen in nova shell models having shell inclination angles between 60-90 $\degr$ \citep{gill}.  \citet{kt} infer that the shell is seen edge-on whereas \citet{eyres} favour a low inclination angle. It is to be noted that similar profiles may be produced by different parameters of inclination angles, ellipticities and positions of rings \citep{gill}. In the absence of detailed modelling, we can only say that the nova spectra reflect the structure  (equatorial ring - polar ring / cap) of the shell.   
As seen in the insets in figures \ref{coronal_apr01} and \ref{coronal_may01}, the nebular line profiles show changes in the red-blue asymmetry during the two epochs, reflecting the fact that shell-shaping is continuing even at this late stage.

\subsection{Ly$\beta$ fluorescence}
A striking feature of the spectra is the presence of strong lines of oxygen. 
The 1.128 $\mu$m line is much stronger than the 1.316 $\mu$m line on all days. Also, the 8446 \AA\, line is stronger than would be expected from recombination alone, which should produce F$_{8446}$/F$_{H\alpha} \le$\, 10$^{-4}$ \citep{rrp}. The 1.128 $\mu$m line is predominantly produced from fluorescent excitation of \mbox{O\,{\sc i}} by Ly$\beta$ (Bowen fluorescence). This is also the dominant mechanism for the 8446 \AA\, line.  The main implication of strong \mbox{O\,{\sc i}}  lines generated by Ly$\beta$ fluorescence is the existence of a region or regions where a sizeable
fraction of the O and H is neutral but where a high Ly$\beta$ flux density
is also present. Such regions are dense, warm, and quite optically thick in the Lyman lines.

\subsection{Unidentified lines}

Several novae have shown  unidentified lines at 0.8926, 1.1110, 1.1900, 1.5545 and 2.0996 $\mu$\,m (for example, V4633 Sgr \citep{lynch01}, V723 Cas \citep{rudy}, CI Aql \citep{lynch04}). They first appear about the same time as the \mbox{He\,{\sc ii}} lines and generally persist till the coronal phase. Thus, they are of medium to high excitation and require relatively low electron densities for appearance \citep{lynch04}. Near-infrared spectra of V1494 Aql display at least two of these lines -- 1.1110 and 1.5545 $\mu$\,m . The Brackett 16 line appears unusually strong in the 31 December 1999 spectrum and it could be blended with the unidentified line at 1.5545 $\mu$\,m. The shoulder on Pa$\gamma$ line on 11 Feb 2000 could be the 1.1110 $\mu$\,m line. There are a few other unidentified lines which are seen only in some novae. For example, the lines at  1.6983 $\mu$\,m (V2487 Oph -- \citealt{lynch00}) and 1.770 $\mu$\,m (V838 Her -- \citealt{hs}). The 1.6983 $\mu$\,m line is seen in V1494 Aql on 31 December 1999 as a blend with the Brackett 11 line, which appears unusually strong. The unidentified line at 1.770 $\mu$\,m  is seen in the 1 March 2000 spectrum.

\subsection{Ejecta ionisation}
\label{ionisation}

The ejecta of V1494 Aql displayed only low excitation lines in the first month following outburst. \mbox{He\,{\sc i}} emission lines were not seen in our December 1999 spectra (only the \mbox{He\,{\sc i}} 1.252 $\mu$\,m line is seen in the 31 December 1999 spectrum), indicating that the level of ionisation was less than 25 eV. The appearance and rapid rise of the  \mbox{He\,{\sc i}} lines in February and March 2000 (see figures \ref{hspec} and \ref{kspec}) suggests that the ejecta was evolving to a higher excitation state in this period. By July 2000, the ejecta ionisation had reached more than 329 eV \citep{venturini}. However, lines of lower ionisation were also seen, suggesting that there were still some denser, cooler zones within the clumpy ejecta. About 17 months after the outburst, such zones hardly existed and the ejecta were almost completely ionised as can be seen in figure \ref{coronal_apr01}. Increased emission line strengths in May 2001 (see figure \ref{coronal_may01}) suggest a further hardening of the nova spectrum.

\subsection{Physical parameters of the ejecta}
\label{abundances}

Emission line fluxes corrected for extinction enable us to determine physical conditions in the ejecta. We have used the nebular stage spectra because they are ideal for abundance estimates.

Since the standard nebular lines used for density estimate are not available in our data, we use the H$\beta$ luminosity to estimate the electron number density, N$_e$. The volume of the line emitting region is estimated assuming the shell to be spherical, with a filling factor of 0.01 (\cite{ie} obtained a value of 0.016 in June 2000) and uniformly expanding with a velocity of 2500 km s$^{-1}$. The electron temperature, T$_e$, in this region is assumed to be 1.5 $\times$ 10$^{4}$ K. Using these simplistic assumptions, the N$_e$ is found to be 1.1 ($\pm$  0.06) $\times$ 10$^5$ cm$^{-3}$. Only the uncertainty in flux is considered for calculating the error. As mentioned in sec. \ref{geometry}, the shell is aspherical and there could exist several velocity components. From the derived N$_e$  we estimate the mass of hydrogen in the ejecta, M$_{\rm H}$, to be 6 $\times$ 10$^{-6}$ M$_{\sun}$. The total ejecta mass would be higher than this value.

Assuming that all the helium is ionised, the helium abundance by number is estimated using the \mbox{He\, {\sc ii}} 4686/H$\beta$ ratio. The hydrogen and \mbox{He\,{\sc ii}} emissivities are from \cite{hummer}. The helium abundance of V1494 Aql is found to be 0.24, and is similar to that observed in other novae such as V1425 Aql \citep{kamath}.

We have used the {\sc nebular} package \citep{shaw} within IRAF to calculate T$_e$ in the zone of nebular lines and ionic abundances. T$_e$ as determined using \mbox{[O\,{\sc iii}]} line fluxes is 1.0 ($\pm$ 0.02) $\times$ 10$^5$ K. The ionic abundances with respect to H$^{+}$ are calculated assming that both the  emission lines -- nebular and H$\beta$ -- arise in regions with the same T$_e$ and N$_e$. This assumption is not strictly true (for example, the O$^{+}$ ion requires T$_e$ $\le$\, 2$\times$ 10$^{4}$ K) but provides a first-order estimate of the abundances in the ejecta. The abundances of O$^{+}$, O$^{2+}$ and S$^{2+}$ are shown in Table \ref{parameters}. As mentioned in Sections \ref{nebular}, \ref{geometry} and \ref{ionisation}, there is a density as well as temperature stratification in the ejecta. In the absence of detailed modelling, 
the values in Table \ref{parameters} are considered to be representative of the conditions in the ejected 
material.

The early spectra show numerous lines of neutral oxygen, carbon and nitrogen. Lines such as \mbox{O\,{\sc i}} 9264~\AA\, and \mbox{N\,{\sc i}} 9042~\AA\ are produced solely by recombination. This indicates an abundance enhancement of these elements. Also, since the mechanism of Ly$\beta$ fluorescence is very strong and persistent for a long time, it indicates the possible presence of a region of fairly high oxygen abundance. Therefore, the CNO abundances seem to be enhanced. This is broadly in line with present nova theories (see \cite{hernanz} for a recent review).

\begin{table}
\caption{Physical conditions in the ejecta during the nebular phase.}
\label{parameters}
\centering
\begin{tabular}{l l l }
\hline\hline
Parameter & Value & Line used  \\
\hline
N$_e$ & 1.1 ($\pm$  0.06) $\times$ 10$^5$ cm$^{-3}$ & H$\beta$ \\
M$_{\rm H}$ & 6 $\times$ 10$^{-6}$ M$_{\sun}$ &   H$\beta$ \\
T$_e$ & 1.0 ($\pm$ 0.02) $\times$ 10$^5$ K & \mbox{[O\,{\sc iii}]} \\
He/H &  0.24 ($\pm$ 0.06) &  \mbox{He\,{\sc ii}} 4686 \AA \\
N(O$^{+}$/N(H$^+$) &  1.45 $\times$ 10$^{-5}$ & \mbox{[O\,{\sc ii}]} 7330 \AA \\
N(O$^{2+}$)/N(H$^+$)  & 1.57 $\times$ 10$^{-6}$ & \mbox{[O\,{\sc iii}]} 4959+5007 \AA \\
N(S$^{2+}$)/N(H$^+$) & 7.5 $\times$ 10$^{-8}$ &  \mbox{[S\,{\sc iii}]} 6312 \AA\\
N(S$^{2+}$)/N(H$^+$) &  3.4 $\times$ 10$^{-7}$ & \mbox{[S\,{\sc iii}]} 9069 \AA\\
\hline
\end{tabular}
\end{table}

\section{Remarks}

We have presented optical and near-infrared spectra of the fast nova V1494 Aquilae 1999 \# 2 in the early decline, transition and nebular phases, covering 18 months since outburst. Based on our data and observations reported in literature, the following picture emerges.

Nova V1494 Aql was a fast nova which ejected matter asymmetrically at velocities of 1000-2500 km s$^{-1}$. The ejected matter was optically thick  initially, but partial thinness set in soon. The ejecta displayed low ionisation levels during the first month after outburst. Higher ionisation lines were evident after about day 65. At about 17 months after outburst, the ejecta were largely ionised and showed strong coronal lines. The clumpy nature of the ejecta was evident in polarisation observations, spectral line profiles, nebular lines and the MERLIN radio image. The nebular spectra present evidence of temperature and density stratification within the ejecta. The calculated elemental and ionic abundances in the ejecta are similar to those found in other novae. 

\section*{Acknowledgments}
We thank all staff members of the respective observatories for help during the observations. Research work at Physical Research Laboratory is funded by the Department of Space, Government of India. IRAF is distributed by the National Optical Astronomy Observatories, which are operated by the Association of Universities for Research in Astronomy, Inc. , under cooperative agreement with the National Science Foundation.

\label{lastpage}


\begin{thebibliography}{99}

\bibitem[\protect\citeauthoryear {Anupama \& Prabhu}{1993}] {anupama} Anupama, G. C. \& Prabhu, T. P., 1993, MNRAS, 263, 335

\bibitem[\protect\citeauthoryear {Anupama, Sahu \& Mayya}{2001}]{gca} Anupama, G.C., Sahu, D. K., \& Mayya Y. D., 2001, BASI, 29, 375

\bibitem[\protect\citeauthoryear {Arkhipova et al.}{2002}]{arkhipova} Arkhipova, V. P., Burlak, M. A., \& Esipov, V. F., 2002, Astron. Letters, 28, 100

\bibitem[\protect\citeauthoryear {Ayani}{1999}]{ayani} Ayani, K., 1999, IAUC 7323

\bibitem[\protect\citeauthoryear {Eyres et al.}{2005}]{eyres} Eyres, S. P. S., Heywood, I., O'Brien, T. J . O. et al., 2005, 358, 1019

\bibitem[\protect\citeauthoryear {Fujii}{1999}]{fujii} Fujii, M., 1999, IAUC 7323

\bibitem[\protect\citeauthoryear {Gill \& O'Brien}{1999}]{gill} Gill, C. D., \& O'Brien, T. J., 1999, MNRAS, 307, 677

\bibitem[\protect\citeauthoryear {Hamuy et al.}{1994}]{hamuy} Hamuy, M, Suntzeff, N. B., Heathcote, S. R. et al., 1994, PASP, 106, 566

\bibitem[\protect\citeauthoryear {Harrison \& Stringfellow}{1994}]{hs} Harrison, T. E., Stringfellow, G. S., 1994, ApJ, 437, 827

\bibitem[\protect\citeauthoryear {Hernanz}{2004}]{hernanz} Hernanz, M., 2004, astro-ph/0412333

\bibitem[\protect\citeauthoryear {Hummer \& Storey}{1987}]{hummer} Hummer, D. G., \& Storey, P. J., 1987, MNRAS, 224, 801


\bibitem[\protect\citeauthoryear {Hutchings}{1972}]{hutchings} Hutchings, J. B., 1972, MNRAS, 158, 177

\bibitem[\protect\citeauthoryear {Ijima \& Esenoglu}{2003}]{ie} Iijima, T., \& Esenoglu, H. H., 2003, A\&A, 404, 997

\bibitem[\protect\citeauthoryear {Kamath et al.}{1997}]{kamath} Kamath, U. S., Anupama, G. C., Ashok, N. M., Chandrasekhar, T., 1997, AJ, 114, 2671

\bibitem[\protect\citeauthoryear {Kawabata et al.}{2001}]{kawabata} Kawabata, K. S., Akitaya, H., Hirakata, N. et al., 2001, ApJ, 552, 782

\bibitem[\protect\citeauthoryear {Kiss \& Thomson}{2000}]{kt} Kiss, L. L., \& Thomson, J. R. 2000, A\&A, 355, L9

\bibitem[\protect\citeauthoryear {Kiyota et al.}{2004}]{kky} Kiyota, S., Kato, T., \& Yamaoka, H., 2004, PASJ, 56, S193

\bibitem[\protect\citeauthoryear {Lynch et al.}{2000}]{lynch00} Lynch, D. K., Rudy, R. J., Mazuk, S.,  Puetter, R. C., 2000, ApJ, 541, 791

\bibitem[\protect\citeauthoryear {Lynch et al.}{2001}]{lynch01} Lynch, D. K., Rudy, R. J., Venturini, C. C., Mazuk, S., 2001, AJ, 122, 2013

\bibitem[\protect\citeauthoryear {Lynch et al.}{2004}]{lynch04} Lynch, D.K., Wilson. J. C., Rudy, R. J., et al., 2004, AJ, 127, 1089

\bibitem[\protect\citeauthoryear ({Moro et al.}{1999}]{moro} Moro, D., Pizzella, A., \& Munari, U., 1999, IAUC 7323 

\bibitem[\protect\citeauthoryear {Pereira}{1999}]{pereira} Pereira, A., 1999, IAUC 7323

\bibitem[\protect\citeauthoryear {Rudy, Rossano \& Puetter}{1989}]{rrp} Rudy, R. J., Rosano, G. S., Puetter, R. C., 1989, ApJ, 346, 799

\bibitem[\protect\citeauthoryear {Rudy et al.}{2002}]{rudy} Rudy, R. J., Venturini, C. C., Lynch, D .K., Mazuk, S., Puetter, R. C., 2002, ApJ, 573, 794

\bibitem[\protect\citeauthoryear {Saizar \& Ferland}{1994}]{saizar} Saizar, P., \& Ferland, G. J., 1994, ApJ, 425, 755

\bibitem[\protect\citeauthoryear {Shaw \& Dufour}{1995}] {shaw} Shaw, R. A., \& Dufour, R. J., 1995, PASP, 107, 896

\bibitem[\protect\citeauthoryear {Starrfield et al.}{2000}]{starrfield} Starrfield, S., Shore, S. N., Butt, Y. et al., 2000, BAAS, 32, 1253

\bibitem[\protect\citeauthoryear {Venturini et al.}{2000}]{venturini} Venturini, et al., 2000, IAUC 7490

\bibitem[\protect\citeauthoryear {Williams}{1992}]{williams} Williams, R. E., 1992, AJ, 104, 725

\bibitem[\protect\citeauthoryear {Williams et al.}{1991}]{ctio1} Williams, R. E., Hamuy, M., Phillips, M. M  et al., 1991, ApJ, 376, 721

\bibitem[\protect\citeauthoryear {Williams et al}{1994}]{ctio2} Williams, R. E., Phillips, M. M., \& Hamuy, M., 1994, ApJS, 90, 297

\end{thebibliography}
\end{document}